\documentclass[preprint,aps,floats,showpacs,nofootinbib]{revtex4}
\usepackage{graphicx}
\usepackage{epsfig}
\usepackage{amsmath}
\usepackage{verbatim}

\def\be{\begin{equation}}
\def\ee{\end{equation}}
\def\bea{\begin{eqnarray}}
\def\eea{\end{eqnarray}}
{\newcommand{\lsim}{\mbox{\raisebox{-.6ex}{~$\stackrel{<}{\sim}$~}}}
{
\def\mpl{M_{\rm {Pl}}}
\def\gev{{\rm \,Ge\kern-0.125em V}}
\def\tev{{\rm \,Te\kern-0.125em V}}
\def\mev{{\rm \,Me\kern-0.125em V}}

\def\z{\zeta}
\def\zk{\zeta_k}
\def\zh{\hat\zeta}
\def\zhat{\hat\zeta}
\def\fnl{f_{\rm NL}}
\def\n{N_*}
\def\Ne{N_e}

\def\t{t}
\def\tinit{t_0}
\def\tex{t_{*}}

\def\eps{\epsilon}
\def\epsex{\epsilon_{*}}
\def\epsilonex{\epsilon_{*}}
\def\etaex{\eta_{*}}

\def\half{\frac12}
\def\phidot{\dot\phi}
\def\tdotphi{\dddot\phi}
\def\fdotphi{\ddddot\phi}
\def\fvdotphi{\frac{d\ddddot\phi}{dt}}
\def\deltaphi{\delta\phi}
\def\partiali{\partial_i}

\def\Q{Q}
\def\Qdot{\dot Q}
\def\phibar{\phi}

\begin{document}
\title{\bf
Remarks on non-gaussian fluctuations of the inflaton and constancy
of $\zeta$ outside the horizon
}
\author{Namit Mahajan}
\email{nmahajan@prl.res.in}
\author{Raghavan Rangarajan}
\email{raghavan@prl.res.in}
\affiliation{Theoretical Physics Division, Physical Research Laboratory,
Navrangpura, Ahmedabad 380 009, India}
%\date{\today}
\begin{abstract}
\noindent
We point out that the non-gaussianity arising from cubic self interactions of
the inflaton field is proportional to $\xi N_e$ where $\xi\sim V'''$ and $N_e$ is 
the number of 
e-foldings from horizon exit till the end of inflation.  
For scales of interest $N_e=60$, and 
for models of inflation
such as new inflation, natural inflation and running mass inflation 
$\xi$ is large compared to the slow roll parameter
$\epsilon\sim V^{\prime2}$.
%, and $N_e=60$ for scales of interest.  
Therefore
the contribution from self interactions should not be outrightly ignored
while retaining other terms in the non-gaussianity parameter $\fnl$.  But the
$N_e$ dependent term seems to imply the growth of non-gaussianities outside the horizon.
Therefore
we briefly discuss the issue of the constancy of correlations of the curvature
perturbation $\zeta$ outside the horizon.
We then calculate the 3-point function of the inflaton fluctuations using the canonical
formalism and further obtain the 3-point function of $\zeta_k$.  We find that the $N_e$ dependent contribution
to $\fnl$ from self interactions of the inflaton field is cancelled by contributions from
other terms associated with non-linearities in cosmological perturbation theory.

\vskip 1cm
\noindent
{\bf Keywords:} Inflationary cosmology, non-gaussianity, curvature perturbation
\end{abstract}
\pacs{98.80.-k,98.80.Cq}
\maketitle
%\section{}
\noindent

Non-gaussianities arising from self interactions of
the inflaton field are
usually treated as negligible compared to those arising from non-linearities
in cosmological perturbation theory \cite{mald,creminelli} because
they are proportional to
higher order slow roll parameters, such as $\xi$.
In this paper we
point out earlier results indicating
that this contribution includes a term proportional
to $\Ne$, the number of e-foldings after the scale of interest has
left the horizon, which for our horizon scale is approximately 60 by the
end of inflation.
Then for certain models of inflation with relatively larger higher order
slow roll parameters, such as new inflation, small field natural inflation
$(f<1.5\mpl)$, and running mass inflation, this contribution can be
comparable to other contributions from non-gaussianities in
cosmological perturbation theory.  

However 
the dependence on $\Ne$ would seem to imply
that $n$-point ($n>2$) correlations of the curvature perturbation $\zeta$ grow
outside the horizon.
Hence we discuss the related issue of constancy of
$\zeta$ outside
the horizon.
The %(classical) 
derivations regarding the constancy
of $\zeta_k$ are classical.  They only 
imply that the 2-point function of $\zh$ to lowest order, $\sim |\zk|^2
\delta^3(\bf{k}-\bf{k'})$,
is constant outside the horizon.
(Hatted quantities denote
quantum operators.)
For higher point functions of $\hat\z$,
there is an additional requirement associated with convergence of an integral over time, as we discuss
below.

Having clarified this we 
then work in a gauge in which
$\delta \phi$, the fluctuation in the inflaton field, is not zero
and recalculate the 3-point function of $\hat{\delta\phi_k}$, using the 
canonical formalism which we believe has not been done before.
This agrees with results of earlier calculations using the path integral approach and
field equations.

We then relate $\hat{\delta \phi_k}$ to $\hat\zeta_k$. 
We argue that the quantisation of
the relation between $\zeta$ and $\delta\phi$ derived from the
$\delta N$ formalism is not ideal if we wish to study any possible growth of 
$n$-point functions of $\hat\zeta$ outside the horizon.  Using a slightly 
different relation which explicitly includes dependence on the final time, 
we then obtain the 3-point function of $\hat\zeta_k$,
including the term associated with self interactions of the
inflaton field which is proportional to $\Ne$.
Our calculation of $\langle \hat\zeta^3\rangle$ is straightforward and does not involve complicated
field redefinitions.
Finally we find that the time dependent self interaction contribution is
cancelled 
by other terms associated with non-gaussianities in cosmological
perturbation theory.

For an inflaton with a cubic interaction the bispectrum parameter
$\fnl$ is 
\begin{eqnarray}
        \frac{6}{5} f_{\rm NL} &=&
 \xi \left[ \frac{1}{3} + \gamma - N_e
+\frac{3} {\sum_i k_i^3}  \left( k_t \sum_{i < j} k_i k_j -
                \frac{4}{9} k_t^3 \right)\right]
\nonumber\\
&&\mbox{} +\frac{3}{2} \epsilon - \eta
+
        \frac{\epsilon}{\sum_i k_i^3} \left(
             \frac{4}{k_t} \sum_{i < j} k_i^2 k_j^2 +
            \frac{1}{2} \sum_{i \neq j} k_i k_j^2
        \right) \,.
        \label{fnl}
    \end{eqnarray}
where $i,j= 1, 2, 3 $, $k_i=|{\bf k}_i|$, $k_t = \sum_i k_i$, 
$N_e$ is the number of e-foldings of inflation from the time the mode of
interest leaves the horizon at $t_{\rm ex}$ till the time $\t$,
which can be at any later time during inflation.
Our expression is similar to that 
in Eq. (38) of Ref. \cite{seeryetal} with
their $t_*$ replaced by $t$, 
and we have replaced $\n$ by
$-N_e = -H(\t-t_{\rm{ex}})$. 
We shall explain the distinction between our expressions later.
The slow variation in $H$ can be ignored
in $N_e$ since $f_{NL}$ is to first order in slow roll.
$\epsilon$, $\eta$ and $\xi$ are the slow roll parameters
evaluated at $\t$.
\begin{eqnarray}
\epsilon 
&\simeq &
        \frac{1}{2} \left( \frac{V'}{V} \right)^2 \simeq
        \frac{1}{2} \frac{\dot{\phi}^2}{H^2}
        \label{epsilon}  \\
        \eta & \equiv & \frac{V''}{V} \simeq
        - \frac{\ddot{\phi}}{H \dot{\phi}} + \epsilon
\label{eta}\\
        \xi & \equiv & \frac{V'V'''}{V^2} 
    \end{eqnarray}
We use $\xi$ rather than $\xi^2$ as
used by some authors and $M_{\rm P}=\mpl/\sqrt {8\pi}$ has been set to 1.
$\gamma \approx 0.577216$
    is Euler's constant.

The contribution due to the cubic interaction of the inflaton is the $\xi$ dependent
term above and was obtained in
Refs. \cite{falketal,
zald,bern,seeryetal}.  The
argument to ignore this term is that it is proportional to $\xi\sim
V^{\prime\prime\prime}$ and is hence negligible compared to terms
proportional to $\epsilon\sim V^{\prime 2}$ and $\eta\sim
V^{\prime\prime}$. This is valid for models such as chaotic inflation for
which $\epsilon>\eta>\xi$. However, in general, the hierarchy is
$\eta>\xi>\sigma>...$, while $\epsilon$ may be larger or smaller than
other slow roll parameters.
In particular, for new inflation, small field natural inflation
$(f<1.5 \mpl)$, and running mass inflation, $\xi>\epsilon$.
In general, for small field models with a concave-downward potential,
$\epsilon\lsim0.0001$ \cite{boublyth}.
Therefore the $\xi$ term should not be outrightly ignored
while retaining other terms in $\fnl$ above.

There is another reason why the $\xi$ term in $\fnl$ is not automatically small
compared to other terms.  For scales of the order of our horizon today
$\Ne$ is about 60 if $\t$ is at the end of inflation.
Therefore the term $\xi \Ne$ in $\fnl$ above can be
large.  For a new inflation potential of the form $V=V_0 -\mu\phi^3\, (\phi>0)$,
$\xi
= 0.5\, \eta^2$.  
Also, $\eta = 0.5 (n_s -1)+3\epsilon\approx
0.5 (n_s -1)$ \cite{alabidilyth}.
If we take
$n_s=0.96$ \cite{wmap} then $\eta$ is -0.02 and $\xi\Ne$ is
0.012.  $\epsilon$ is much smaller.  Thus we see that $\xi\Ne$ is
comparable to $\eta$ and $\epsilon\ll \xi\Ne$. 
%$n_s=0.96$ \cite{wmap} then $\eta_*$ is -0.02 and $\xi_*\Ne$ is
%0.012.  $\epsilon_*$ is much smaller.  Thus we see that $\xi_*\Ne$ is
%comparable to $\eta_*$ and $\epsilon_*\ll \eta_*$. 
\footnote{Interestingly, Appendix B of Ref. \cite{barnabycline}
obtains a similar result but erroneously concludes that the $\Ne$ dependent
result
of Ref. \cite{falketal} and the $\Ne$ independent result of Ref. \cite{mald}
are the same because they are of the same order.}
\footnote{In Ref. \cite{seeryetal} it is argued that evaluating
expectation values at the end of inflation may not be valid 
for large $\Ne\approx60$ because of 
divergences of the form $\epsilon^{m+2} \Ne^m \,(m\ge1)$.
However for the potential we are considering
$\epsilon$ is much smaller than 1/60.}

The
presence of the time dependent $N_e$ in $\fnl$
seems to contradict the notion that $n$-point functions
of $\hat\z$ do not grow outside the horizon.  
It may be argued that since the curvature perturbation $\z$
does not grow outside the horizon the $\Ne$ contribution from the
non-gaussianities of the inflaton can not contribute to $\fnl$. Note,
however, that this term has been obtained independently in Refs.
\cite{falketal,
zald,bern,seeryetal}.  So we now consider the literature on the
constancy of $\z$ in the context of non-linear cosmological
perturbation theory.  

Salopek and Bond \cite{SB} first introduced the
generalisation of the Bardeen-Steinhardt-Turner variable $\z$ \cite{BST}
for the case when one wishes to consider 
non-gaussianities in the curvature perturbation.
The Salopek-Bond $\z(x)$ is constant outside the horizon. 
In other works,
such as
Refs. \cite{lythwands,rigopshell,malikwands,lythmaliksasaki,
lv1,lv2},
the constancy of $\z$ is
shown for a %[classical] 
$\z_k$ mode.
The above works deal with classical $\z$ and the constancy of 
classical $\z(x)$ or $\zk$
only implies that 
the quantum 2-point function
$\langle\zh_{\bf k_1}\zh_{\bf k_2}\rangle
=(2\pi)^3|\zeta_{k_1}|^2\delta(\bf{k_1}-\bf{k_2})$ is constant outside the horizon
(at lowest order).
\footnote{Higher
order corrections from loops can give time
dependent or $\Ne$ dependent contributions
as mentioned in Sec. VI of Ref. \cite{weinbergII}.}
But
it is not obvious that
%this will be equivalent to asking whether
%$\langle \z \z\rangle$ is constant outside the horizon or
%all $n(>2)$-point 
other higher $n$-point
functions of
$\hat\z$ are constant outside the horizon.

One may argue that after horizon exit the curvature perturbations are classical
and so the constancy of $\zeta$ should imply constancy of $n$-point
function for $n>2$ also.  As argued by Weinberg in Ref. \cite{weinbergII}, the
perturbations are classical in the sense that commutators involving $\hat\zeta$
and its time derivative go to 0 
for large
$t$.  However, this implies that quantum $n$-point functions of zeta can nevertheless grow outside the horizon,
but only as powers of $\ln a$ and not as powers of $a$.   It is the time dependence 
due to $\ln a \sim N_e$ for the 3-point function which is the focus of this
article. 

The $n$-point functions of $\hat\zeta$ are given by \cite{weinbergII,mald}
\begin{eqnarray}
\langle \hat O(t)\rangle &=&\sum_{N=0}^\infty i^N\, \int_{\tinit}^t dt_N
\int_{\tinit}^{t_N} dt_{N-1} \cdots \int_{\tinit}^{t_2} dt_1 \nonumber\\
&&\times\left\langle \Big[\hat H_I(t_1),\Big[\hat H_I(t_2),\cdots
\Big[\hat H_I(t_N),\hat O_I(t)\Big]\cdots\Big]\Big]\right\rangle\;,
\end{eqnarray}
where $\hat O(t)$ can be any product of $\zh$ operators, $\hat O_I(t)$ is
$\hat O(t)$ in the interaction picture generated by the quadratic
part of the Hamiltonian, and $t_0$ is some
early time.  Note that the expectation values are obtained in the
in-in formalism and so the bra-s and kets refer to
$_{in}\langle 0|$ and $|0\rangle_{in}$ respectively.
$\hat H_I$ is the interaction Hamiltonian and includes
terms that are third or higher order in $\zh$.
For the three-point function at lowest order this reduces to
\be
\langle \hat\zeta^3(t) \rangle = i \int_{t_0}^t dt'\Big\langle \Big[
\hat H_{I}(t'),
\hat\zeta_I^{3}(t)
\Big] \Big\rangle
\label{zeta3}
\ee
If $\zeta_k$ is constant,
$\hat\zeta \sim e^{i{\bf k.x}} \zeta_k c_k + e^{-i{\bf k.x}}\zeta_k^* c_k^\dagger$ is constant.
But for $\langle\zh^3(t)\rangle$ to be constant outside
the horizon
one must ensure that the contribution to the integral above
from $t_{\rm ex}$ to $t$ is suppressed.
Note that
$\hat\zeta$ is related to $\hat{\delta\phi}$ and
$\langle \hat{(\delta\phi)}^n\rangle$ grows outside the horizon.  

The convergence of the integral
for large $t$, and certain other conditions for the constancy of 
$\langle{\hat{\zeta}}^n\rangle$ outside the horizon have been discussed in general 
in Ref. \cite{weinbergIII}.  
But in Ref. \cite{weinbergIII} (see Eq. (29))
only gaussian fluctuations of the inflaton are considered.
\footnote
{
Note that Ref. \cite{SB} also considers only
gaussian fluctuations of the inflaton, since 
$\delta\phi$ is set equal to $H/(2\pi)$ in the evaluation of $\z$
in Secs. IIIB and IIID. 
}  
So one
should
verify whether or not $\langle{\hat{\zeta}}^n\rangle$ is indeed constant outside
the horizon when one includes non-gaussian fluctuations of the inflaton.
We now check this explicitly for the three point 
function of $\hat\zeta$ while including a cubic interaction of the inflaton field.
The 3-point function of $\hat\zeta$ has been obtained by other authors.  Largely,
the self-interactions of the inflaton are ignored.  Moreover, assuming that
there is no contribution outside the horzion, the integral
in Eq. (\ref{zeta3}) is cut off at $\tex$.  In cases where one first 
relates $\hat\zeta$ to $\hat\delta\phi$ using the $\delta N$ formalism
and then calculates the 3-point function,
obtaining any contribution from evolution 
outside the horizon is precluded by the adopted formalism. 

We work
in a gauge in which
$\delta\phi \neq 0$.  We first calculate $\langle\hat{(\delta\phi)}^3\rangle$
using the equivalent of Eq.~(\ref{zeta3}) for the 3-point function of
$\hat{\delta\phi}$.  Our results agrees with those obtained
in Ref. \cite{seeryetal} using field equations.
We then relate
$\zeta$ to $\delta\phi$ and use this
to obtain $\langle\hat\zeta^3\rangle$ and $\fnl$.
The $\Ne$ dependent
term mentioned
above associated with the cubic self interaction of the inflaton does
appear in $\fnl$.
The 3-point function $\langle\hat\zeta^3\rangle$ has also been obtained in a
gauge in which $\delta\phi$ is 0 \cite{mald,chenetal}.
Since these calculations do not include self-interactions of the inflaton their
results do not include the $N_e$ term.

The relevant part of the action $S$ for $\delta\phi$ can be expressed 
as the sum of terms quadratic and
cubic in $\deltaphi$.  For notational convenience we hereafter replace $\deltaphi$
with $\Q$, and let $\phi$ represent the background homogeneous field.
Then
\be
S=S_2+S_3
\ee
where \cite{seerylidsey,sasakistewart} 
\be
S_2=\int\, dt\, d^3x\, a^3 \left[\half (\dot \Q)^2 - \frac{1}{2a^2} (\partiali \Q)^2
-\half\left\{
V^{''}(\phi) -\frac{1}{a^3} \frac{d}{dt}
\left(\frac{a^3}{H}\dot\phi^2\right)
\right\} \Q^2
\right]\,,
\ee
and $S_3$ is given in Eq. (3.6) of Ref. \cite{mald}.
Retaining terms to
leading order in slow roll parameters we obtain
\be
S_3=\int\, dt\, d^3 x\, a^3
\left[
-\frac{\phidot}{4H} \Q \Qdot^2
-\frac{1}{a^2} \frac{\phidot}{4H} \Q (\partial_i \Q)^2
-\frac{1}{a^2}
\partial_i \Q\partial_i\psi\Qdot 
%\Qdot \partial_i\psi\partial_i \Q
-\frac{1}{6}V'''(\phi) \Q^3 \right]
\ee
where
\be
\partial^2\psi=-\frac{a^2}{2H} \phidot \Qdot
\label{psi}
\ee
In Ref. \cite{mald}
the last term in $S_3$ above, which is
proportional to $\xi$, was ignored.  As we have
argued earlier the contribution of
this term could actually be larger than that of other terms
above for certain models of inflation.  Therefore it ought not to be ignored
at this juncture.

The shift function $N_i=\partial_i\psi$, and $\psi$ differs from
$\chi$ of Ref. \cite{mald} by a factor of $a^2$.
Using eq. (\ref{psi}), $S_3$ can be rewritten as
\footnote{
The action $S_3$ provided in Ref. \cite{seerylidsey} may contain
typographical errors.  The prefactor for the last term of
$S_3$ in Eq. (53) of Ref. \cite{seerylidsey} should be
$a^{-2}$ rather than $a^{-4}$, and the definition
of $\psi$ and $N_i$ in Eq. (43) is not in agreement
with eq. (2.24) of Ref. \cite{mald}.  However, Eq. (54) for
$\psi$ is correct.
}
\be
S_3=\int\, dt\, d^3 x\, a^3
\left[
-\frac{\phidot}{4H} \Q \Qdot^2
-\frac{1}{a^2} \frac{\phidot}{4H} \Q (\partial_i \Q)^2
+\frac{\phidot}{2H}
\partial_i \Q(\partial_i^{-1}\Qdot)\Qdot 
-\frac{1}{6}V'''(\phi) \Q^3 \right]
\ee

One can obtain $H$, and thus $H_I$, from the lagrangian in $S_2+S_3$.
We follow Ref. \cite{IZ} for dealing with the
$\Qdot$ dependent interaction terms.  
After obtaining $H_I(\Q,\Pi_Q)$ we
replace $Q$ and $\Pi_Q$ by $Q_{in}$ and $\Pi_{in}$ respectively, and then set
$\Pi_{in}= a^3\,\Qdot_{in}$.
Keeping terms upto first order in $\phidot/H$, as in the action,
we then get $H_I(\Q,\Qdot)=-L_{int}$.  Hereafter we drop the subscript
$in$.

The contribution to $\langle {\hat{\Q}}(\vec{k}_1,t)
\hat{\Q}(\vec{k}_2,t) \hat{\Q}(\vec{k}_3,t) \rangle$
from each term in $H_I$ is given below.  We use the
conformal time $\tau$, defined by $dt=a\,d\tau$, 
%$\tau=-1/(aH)$ 
instead of $t$ and let
the initial time  correspond to $\tau=-\infty$.
The final time corresponds to the reheat time.
\\

1. The $\Q\Qdot^2$ term
\bea
I_1 &=& (-i) \Q_{k1}(\tau) \Q_{k2}(\tau) \Q_{k3}(\tau)
(2\pi)^3\delta^3(\sum_i \vec{k_i})
\int_{-\infty}^{\tau} \, d\tau'
a^2 \frac{\phidot}{4H}
\left[\Q_{k1}^*(\tau') \frac{d\Q_{k2}^*(\tau')}{d\tau'}
\frac{d\Q_{k3}^*(\tau')}{d\tau'} + \rm{perm}\right] \nonumber \\ &+& \rm{c.c.}
\eea
$\Q_k$ is given by \cite{LindeBook}
\bea
\Q_k&=&\frac{i H}{k\sqrt{2k}}
\left(1-i\frac{k}{aH}\right)
\exp\left(i\frac{k}{aH}\right)\cr
&=&\frac{i H}{k\sqrt{2k}}
\left(1+{k}{\tau}\right)
\exp\left(-i{k}{\tau}\right)
\eea
which reduces to $i H/({k\sqrt{2k}})$ for
$|k\tau|\ll 1$.  
\footnote{
There are higher order corrections to the late time mode functions, as discussed
in Ref. \cite{chenetal}.
}
There are 6 permutations of
${k_1,k_2,k_3}$ for the expression within
the integral.  Then
\be
I_1=-\frac{i}{4}\frac{H^3(\tau)}{\prod_i (2k_i^3)}
(2\pi)^3\delta^3(\sum_i \vec{k_i})
\int_{-\infty}^{\tau} \, d\tau'
\phidot(\tau')
\left[k_2^2 k_3^2\,(1-ik_1\tau') e^{ik_t\tau'}
+ \rm{perm}\right]  + \,\,\rm{c.c.}
\ee
where $k_t=k_1+k_2+k_3$.
Replacing the lower limit, $-\infty$, by $-\infty (1- i\delta)$
and setting $\delta$ to 0 after taking the limit eliminates
the contribution of the lower limit. Integrating by parts,
taking the limit $k_i\tau\ll1$, and 
using the complex conjugate to avoid listing some
terms, we get
\begin{eqnarray}
I_1&=&-\frac{i}{4}\frac{H^3(\tau)}{\prod_i (2k_i^3)}
(2\pi)^3\delta^3(\sum_i \vec{k_i})\nonumber\\
&&
\left
[
k_2^2 k_3^2
\left\{
\frac{\phidot(\tau)}{ik_t}
+
\frac{i k_1\phidot(\tau)}{(ik_t)^2}
+
\frac{i k_1\tau\phidot^\prime(\tau)}{(ik_t)^2}
+
\frac{\phidot^{\prime\prime}(\tau)}{(ik_t)^3}
\right.
\right.
\nonumber\\
&&
+
\frac{3i k_1 \phidot^{\prime\prime}(\tau)}{(ik_t)^4}
+
\frac{i k_1 \tau \phidot^{\prime\prime\prime}(\tau)}{(ik_t)^4}
+
\frac{\phidot^{\prime\prime\prime\prime}(\tau)}{(ik_t)^5}
\nonumber\\
&&
\left.\left.
-
\int_{-\infty}^\tau
\, d\tau'
[-5i k_1 \phidot^{\prime\prime\prime\prime}(\tau')
+(1-ik_1\tau')\phidot^{\prime\prime\prime\prime\prime}(\tau')]
\frac{e^{ik_t\tau'}}{(ik_t)^5}
\right\}
+ \rm{perm}
\right] + \,\,\rm{c.c.}
\label{I1_1}
\end{eqnarray}
where $\phidot^\prime=d\phidot/d\tau$ and so on.

In the Appendix we assess the higher derivative terms.  We find that
the $\dddot\phi$ 
term is proportional to
$\eta^2 e^{2N_e}$.  Similarly the
$\fdotphi$ term is proportional to
$\eta^4 e^{4N_e}$.  These higher order terms in
slow roll parameters are
(increasingly) larger than the terms proportional
to $\phidot$ (for $\eta=0.02$ and $N_e=60$).  However it is
not consistent to consider them here as we have ignored terms
higher order in slow roll parameters in the action.  But this
is an indication that there may be convergence issues at higher
orders in the slow roll parameters and these will have to be
handled with care.  
\footnote{
A concern regarding using perturbation theory in
slow roll parameters may also be found in Ref. \cite{seeryetal}, as mentioned earlier.
}
Similar behaviour may be expected while working
in the $\delta\phi=0$ gauge
as, for example, in Ref. \cite{chenetal}
where 
integrals for $\langle\zh^3\rangle$
include powers of
$\epsilon\simeq 0.5 \,\phidot^2/H^2$ 
in the integrand.

Having noted our concern above, we hereafter do not include
terms higher order in slow roll parameters.
Then
\be
I_1=-2 \times \frac{1}{4}\frac{H^3(\tau)}{\prod_i (2k_i^3)}
(2\pi)^3\delta^3(\sum_i \vec{k_i})
\phidot(\tau)
\left[
\frac{k_2^2 k_3^2}{k_t}
+\frac{k_1k_2^2 k_3^2}{k_t^2}
+ \rm{perm}\right]
%+ \,\,\rm{c.c.}
\ee
The prefactor  of 2 comes from the complex conjugate.
%The complex conjugate will give a factor of 2.
There are a total of 6 permutations of the variables $(k_1,k_2,k_3)$.
The interchange
of $k_2$ and $k_3$ gives the same expression as above.
\\

2. The $\Q (\partial_i \Q)^2$ term
\bea
I_2&=&(-i) \Q_{k1}(\tau) \Q_{k2}(\tau) \Q_{k3}(\tau)
(2\pi)^3\delta^3(\sum_i \vec{k_i})\cr
&&
\int_{-\infty}^{\tau} \, d\tau'
 a^2\frac{\phidot}{4H}
[
%\frac
{(-\vec{k_2}\cdot\vec{k_3})}\,
%{\tau'^2}
\Q_{k1}^*(\tau') \Q_{k2}^*(\tau')
\Q_{k3}^*(\tau') + \rm{perm}] + \,\,\rm{c.c.}\cr
&=&-\frac{i}{4}\frac{H^3(\tau)}{\prod_i (2k_i^3)}
(2\pi)^3\delta^3(\sum_i \vec{k_i})\cr
&&\int_{-\infty}^{\tau} \, d\tau'
\phidot(\tau')
\left[
(\frac{-\vec{k_2}\cdot\vec{k_3}}{\tau'^2})
(1-ik_1\tau')(1-ik_2\tau')(1-ik_3\tau')
e^{ik_t\tau'}
+ \rm{perm}\right]  + \,\,\rm{c.c.} \cr
&=&
-2\frac{1}{4}\frac{H^3(\tau)}{\prod_i (2k_i^3)}
(2\pi)^3\delta^3(\sum_i \vec{k_i})
\phidot(\tau)
(\vec{k_2}\cdot\vec{k_3})
\left[ -k_t +
\sum_{i\ne j}\frac{k_i k_j}{k_t}
+\frac{k_1 k_2 k_3}{k_t^2}
+ \rm{perm}\right]
\eea
Clearly there is a symmetry in $(\vec{k_2},\vec{k_3})$ interchange.
$\vec{k_2}\cdot\vec{k_3}$ can be replaced by $(k_1^2-k_2^2-k_3^2)/2$
using $\sum \vec{k_i}=0$.  Note that the first term above is obtained
from the real part of $i\exp[i k_t \tau]/\tau$ in the limit $k_t\tau\ll1$.
\\

3. The $\partial_i\Q (\partial_i^{-1}\Qdot) \Qdot$ term
\bea
I_3&=&(+i) \Q_{k1}(\tau) \Q_{k2}(\tau) \Q_{k3}(\tau)
(2\pi)^3\delta^3(\sum_i \vec{k_i})\cr
&&
\int_{-\infty}^{\tau} \, d\tau'
a^2 \frac{\phidot}{2H}
\frac{ \vec{k_1}\cdot\vec{k_2} } {k_2^2}
[\Q_{k1}^*(\tau') \frac{d\Q_{k2}^*(\tau')}{d\tau'}
\frac{d\Q_{k3}^*(\tau')}{d\tau'} + \rm{perm}] + \,\,\rm{c.c.}\cr
&=&2\frac{1}{2}\frac{H^3(\tau)}{\prod_i (2k_i^3)}
(2\pi)^3\delta^3(\sum_i \vec{k_i})
\phidot(\tau)
\frac{ \vec{k_1}\cdot\vec{k_2} }{ k_2^2}
\left[
\frac{k_2^2 k_3^2}{k_t}
+\frac{k_1k_2^2 k_3^2}{k_t^2}
+ \rm{perm}\right]\cr
&=&2\frac{1}{2}\frac{H^3(\tau)}{\prod_i (2k_i^3)}
(2\pi)^3\delta^3(\sum_i \vec{k_i})
\phidot(\tau)
(\vec{k_1}\cdot\vec{k_2})
\left[
\frac{ k_3^2}{k_t}
+\frac{k_1 k_3^2}{k_t^2}
+ \rm{perm}\right]
%+ \,\,\rm{c.c.}
\eea

The 3-point function of $\hat\Q$ is $I_1+I_2+I_3$ plus the contribution
from the cubic self-interaction.  For the self-interaction
contribution we use the expression given
in Ref. \cite{seeryetal} which agrees with Refs. \cite{falketal,bern}.
Using
Mathematica to include all the permutations and then simplify
their sum gives  
    \begin{eqnarray}
&&
        \langle {\hat{\Q}}(\vec{k}_1,t)
\hat{\Q}(\vec{k}_2,t)
        \hat{\Q}(\vec{k}_3,t) \rangle
%       & \supseteq &
%%%&=&
\nonumber\\&&
=
        (2\pi)^3 \delta(\vec{k}_1
+ \vec{k}_2 + \vec{k}_3)
%\left[
\Biggr[
        \frac{H^2 V'''}{4 \prod_i k_i^3}
%%%        \times
%\right.
%%%\nonumber\\ & &
%%%        \quad
        \left(
            - \frac{4}{9} k_t^3 + k_t \sum_{i < j} k_i k_j +
            \frac{1}{3} \Big\{ \frac{1}{3} + \gamma +
                \ln | k_t \tau | \Big\} \sum_i k_i^3
        \right)\nonumber\\
        & & \quad
\quad\quad\quad\quad\quad\quad
%\left.
        +\,\frac{H^4}{8 \prod_i k_i^3} \frac{\dot{\phi}}{H}
\frac{1}{k_t}
        \left(
            \frac{1}{2} \sum_i k_i^4 -
            5 \sum_{i < j} k_i^2 k_j^2 -
            \sum_{i \neq j\neq k} k_i^2 k_j k_k
        \right)
%\right]
\Biggr]\\
&=&
        (2\pi)^3 \delta(\vec{k}_1
+ \vec{k}_2 + \vec{k}_3)
%\left[
\Biggr[
        \frac{H^2 V'''}{4 \prod_i k_i^3} \times
%\right.
\nonumber\\ & &
        \quad \left(
            - \frac{4}{9} k_t^3 + k_t \sum_{i < j} k_i k_j +
            \frac{1}{3} \Big\{ \frac{1}{3} + \gamma +
                \ln | k_t \tau | \Big\} \sum_i k_i^3
        \right)\nonumber\\
        & & \quad
%\left.
        +\frac{H^4}{8 \prod_i k_i^3} \frac{\dot{\phi}}{H}
        \left(
            \frac{1}{2} \sum_i k_i^3 -
            \frac{4}{k_t} \sum_{i < j} k_i^2 k_j^2 -
            \frac{1}{2} \sum_{i \neq j} k_i k_j^2
        \right)
%\right]
\Biggr]\,,
\label{Q3}
    \end{eqnarray}
where $i,j= 1, 2, 3 $, $k_t = \sum_i k_i$, and we take
all $\vec k_i$ have approximately the same magnitude.  
$\tau=-1/(aH)$.
The term proportional to $V^{\prime\prime\prime}$, which is due
to the cubic self-interaction of the inflaton, was obtained in
Refs. \cite{falketal,
%ganguietal,
zald,bern,seeryetal}.  
$\ln |k_t\tau|=\ln [(aH)_{{\rm ex}}/ (a H)]\approx \ln[a_{{\rm ex}}/a]=
-N_e$, and thus one gets an $N_e$ dependent term.
\footnote{
The result in Ref. \cite{zald} differs by a sign and a term
\cite{seeryetal}.  In addition, $\ln | k_t \eta |$ 
in Ref. \cite{zald} should
be set to $-N_e$ and not $+N_e$.
Ref. \cite{ganguietal} also obtains a $V^{\prime\prime\prime}$
term with a $\ln a$ dependence.
}
This term is not explicitly cancelled by any other term in the expression 
above.  
Moreover,
as we show later, this term is also not cancelled by the time variation
of slow roll parameters in other terms.

The above uses the action/Lagrangian and the canonical formalism
to calculate the 3-point function of $\hat\Q$.
Ref. \cite{seerylidsey} uses the path integral formalism to obtain
the 3-point function.  (It does not consider the contribution from
the cubic self interaction term.)
Ref. \cite{seeryetal} uses the 
solutions of the Heisenberg field equations to obtain
$\langle\hat{\Q}^3\rangle$, including the self interaction
contribution, and our results agree with Eqs.
(20) and (29) of Ref. \cite{seeryetal}.
Ref. \cite{vernizziwands} explicitly
shows the equivalence of the form of $\langle\hat\Q^3\rangle$ obtained in
Refs. \cite{seerylidsey,seeryetal}.

In the $\delta N$ formalism,
the gauge invariant quantity $\zeta(\vec{x},\t)$ is 
the difference in the number of e-foldings of evolution
between some time $\tex$ and $\t$ at $\vec x$ and the 
%(spatial) average
number of e-foldings between $\tex$ and $\t$ for an isotropic
homogeneous background, where $\tex$ lies on 
a spatially flat slice of spacetime with field values 
$\phi(\vec{x},\tex)$ while $\t$ belongs to a spacetime slice
of uniform energy density.
$\tex$ is typically chosen to be a few e-foldings after the relevant
scale has left the horizon.
\bea
\zeta(\vec{x},\t)
&=&
N[\rho(\t),\phi(\vec{x},\tex)]-
N[\rho(\t),\phibar(\tex)] \nonumber \\ \nonumber \\
&=&
\frac
{\partial N[\rho(\t),\phi(\vec{x},\tex)]}
{\partial \phi(\vec{x},\tex) }
\biggr|_{\phibar(\tex)}
\delta\phi(\vec{x},\tex) \nonumber \\ \nonumber \\
&&+\half\,
\frac
{\partial^2 N[\rho(\t),\phi(\vec{x},\tex)]}
{\partial \phi(\vec{x},\tex)^2 }
\biggr|_{\phibar(\tex)}
\delta\phi(\vec{x},\tex)^2
+\cdots
\label{zetaxtaylor}
\eea
where $\phibar(\tex)$ is the spatial average value of $\phi$ at $\tex$,
$\delta\phi(\vec{x},\tex)=\phi(\vec{x},\tex)-\phibar(\tex)
=\Q(\vec{x},\tex)$, and $\cdots$ refers to higher order terms that have 
been omitted. (We have temporarily reintroduced $\phi(\vec{x},t)$).
Dependence of $N$ on $\dot\phi(t_*)$ is ignored in the slow roll approximation
\cite{sasakistewart}.
$N[\rho(\t),\phi(\vec{x},\tex)]$ is given by
\bea
N[\rho(\t),\phi(\vec{x},\tex)]
&=&\int_{\tex}^{\t} H[\phi(\vec{x},t)]\, dt \nonumber \\
&=&
\int_{\phi(\vec{x},\tex)}^{\phi(\t)} H[\phi(\vec{x},t)]\, 
\frac{d\phi(\vec{x},t)}{\phidot(\vec{x},t)}
\eea
Then 
\bea
\frac
{\partial N[\rho(\t),\phi(\vec{x},\tex)]}
{\partial \phi(\vec{x},\tex) }
\biggr|_{\phibar(\tex)}
&=&
-\frac{H[\phibar(\tex)]}{\dot\phibar(\tex)}\cr
&\simeq&
\frac{V}{V'}
\biggr|_{\phibar(\tex)}
\simeq\pm\frac{1}{\sqrt{2\epsex}}
\label{Nd1}\\
\frac
{\partial^2 N[\rho(\t),\phi(\vec{x},\tex)]}
{\partial \phi(\vec{x},\tex)^2 }
\biggr|_{\phibar(\tex)}
&\simeq&
\frac{\partial}{\partial\phi}
\left(
\frac{V(\phi)}{V'(\phi)}
\right)
\biggr|_{\phibar(\tex)}
\simeq
1-\frac{\etaex}{2\epsex}
\label{Nd2}
\eea
The $+\,(-)$ in the first equation is for $V'(\phi) > 0$ ($<0$), or
equivalently, for $\phidot< 0$ ($>0$).  Below we will consider the sign as
for $V'(\phi)<0$ as relevant for new inflation.
(However the various contributions to $\langle\hat\zeta^3\rangle$ are
ultimately dependent on only even powers of $\phidot$,
and so the final expression for
$\langle\hat\zeta^3\rangle$, or $\fnl$,
is independent of the sign of $V'(\phi)$.)
Then $\zeta$ is related to ${Q}$ by
    \begin{equation}
%        \hat
\zeta(\vec{k},\t) = \pm\frac{1}{\sqrt{2 \epsex}}
%        \hat
{Q}(\vec{k},\tex)
        + \frac{1}{2}
\left(
1 - \frac{\etaex}{2\epsex}
\right)
        \int \frac{d^3 q}{(2\pi)^3}
\;
%        \hat
{Q}(\vec{k}_1 - \vec{q},\tex)
%\hat
{Q}(\vec{q},\tex) +
        \cdots ,
        \label{zetaktaylor}
    \end{equation}
The $+\,(-)$ in front of the first term is for $V'(\phi) > 0$ ($<0$), or
equivalently, for $\phidot< 0$ ($>0$).  

In the $\delta N$ formalism, $\zeta$ is independent of $\t$, and there
is no $\t$ dependence on the r.h.s. of Eq. (\ref{zetaktaylor}).
If one directly quantises the above relation, 
as in Refs. \cite{seeryetal,seerylidsey},
then
Eq. (\ref{zetaktaylor}) 
implies
\bea
&&\langle \zh(\vec{k}_1,\t) \zh(\vec{k}_2,\t)
\zh(\vec{k}_3,\t) \rangle\cr
&&=
-\frac{1}{(2\epsilonex)^\frac{3}{2}}
\langle {\hat{\Q}}(\vec{k}_1,\tex)
\hat{\Q}(\vec{k}_2,\tex)
        \hat{\Q}(\vec{k}_3,\tex) \rangle
        \cr
&&
+\frac{1}{2\epsilonex}
\half\left(1-\frac{\etaex}{2\epsilonex}\right)
\langle
{\hat{\Q}}(\vec{k}_1,\tex)
\hat{\Q}(\vec{k}_2,\tex)
\int\, \frac{d^3q}{2\pi^3}
\hat\Q(\vec{k}_3-\vec{q},\tex)\Q(\vec{q},\tex)
+ \rm{perm} \,\rangle
\eea
The r.h.s. above does not depend on $\t$ but that is because of the
way $\hat\zeta(\t)$ was defined as in terms of $\hat Q(\tex)$.

One may instead argue that
the quantum field $\hat\zeta(\t)$ should be expressed as a function of 
quantum operators at $\t$.
Furthermore,
$\t$ and $\tex$ are defined on different hypersurfaces and so correspond to
different `definitions' of time.  
We need a relation between the quantum operators
$\hat\zeta(t)$ and $\hat Q(t)$ which is valid at all times $t$, including 
prior to and after horizon exit, and both operators should be functions of the
same $t$.  For this one may use Eq. (A8) of Ref. \cite{mald} and
    \begin{equation}
        \hat
\zeta(\vec{k},t) = \pm\frac{1}{\sqrt{2 \eps}}
        \hat
{Q}(\vec{k},t)
        + \frac{1}{2}
\left(
1 - \frac{\eta}{2\eps}
\right)
        \int \frac{d^3 q}{(2\pi)^3}
\;
        \hat
{Q}(\vec{k}_1 - \vec{q},t)
\hat
{Q}(\vec{q},t) +
        \cdots ,
        \label{zetaktaylor_mald}
    \end{equation}
`$\cdots$' includes terms quadratic in $\hat\Q$ but which will not contribute
to the three point function because their coefficients are suppressed for 
$t>t_{\rm{ex}}$, i.e., outside the horizon.  
(As discussed in Sec. 3 of Ref. \cite{mald},
$\zeta$ and $\delta\phi$ are defined in different gauges, i.e, 
they are functions
of different time variables $t$ (uniform density gauge) and 
$\tilde t$ (spatially flat or uniform curvature gauge) 
respectively.  But 
$\tilde t = t + T(t,\vec {x})$ and so $\hat Q$ on the r.h.s. of the equation 
above can be expressed in terms of the 
same time variable as on the l.h.s.)
Then
\bea
&&\langle \zh(\vec{k}_1,t,) \zh(\vec{k}_2,t)
\zh(\vec{k}_3,t) \rangle\cr
&&=
-\frac{1}{(2\epsilon)^\frac{3}{2}}
\langle {\hat{\Q}}(\vec{k}_1,t)
\hat{\Q}(\vec{k}_2,t)
        \hat{\Q}(\vec{k}_3,t) \rangle
        \cr
&&
+\frac{1}{2\epsilon}
\half\left(1-\frac{\eta}{2\epsilon}\right)
\langle
{\hat{\Q}}(\vec{k}_1,t)
\hat{\Q}(\vec{k}_2,t)
\int\, \frac{d^3q}{2\pi^3}
\hat\Q(\vec{k}_3-\vec{q},t)\hat\Q(\vec{q},t)
+ \rm{perm} \,\rangle
\label{zeta3ptfunc}
\eea
The first term can be evaluated using Eq. (\ref{Q3}).
The expectation value in the second term above is
\bea
I_4&=&
(2\pi)^3\delta^3(\sum_i \vec{k_i})
\left[ 2\Q_{k1}(t)\Q_{k1}^*(t)\Q_{k2}(t)\Q_{k2}^*(t)
+\rm{perm}\right]\\
&=&8
%\epsilon_\ast^2
\epsilon^2
(2\pi)^3\delta^3(\sum_i \vec{k_i})
[P_{\zeta}(k_1) P_\zeta(k_2) + \rm{perm}]
\eea
The factor of 2 on the first line above
comes from different ways of contracting
the $\hat\Q$s, and there are 3 permutations involving
$k_1,k_2,k_3$.
$P_\zeta$ is defined as
\be
P_\zeta(k)=\frac{1}{2\epsilon}\frac{H^2}{2k^3}
\ee
\begin{comment}
The expectation value in the second term above is
\bea
I_4&=&
(2\pi)^3\delta^3(\sum_i \vec{k_i})
\left[ 2\Q_{k1}(\tex)\Q_{k1}^*(\tex)\Q_{k2}(\tex)\Q_{k2}^*(\tex)
+\rm{perm}\right]\\
&=&8
%%\epsilon_\ast^2
\epsilonex^2
(2\pi)^3\delta^3(\sum_i \vec{k_i})
[P_{\zeta}(k_1) P_\zeta(k_2) + \rm{perm}]
\eea
The factor of 2 on the first line above
comes from different ways of contracting
the $\hat\Q$s, and there are 3 permutations involving
$k_1,k_2,k_3$.
$P_\zeta$ is defined as
\end{comment}
The terms dropped in Eq. (\ref{zetaktaylor_mald}) would have
contributed similar to the second term in Eq. (\ref{zeta3ptfunc}),
i.e., without any time integral as for the first term, 
and hence would be evaluated
only at a time $t>t_{\rm{ex}}$, when their contribution will be suppressed
as mentioned above. 
Defining $\fnl$ 
as in 
Ref. \cite{seeryetal}
    \begin{equation}
        \langle \zh(\vec{k}_1,\t) \zh(\vec{k}_2,\t)
\zh(\vec{k}_3,\t) \rangle
        \equiv (2\pi)^3 \delta(\vec{k}_1 + \vec{k}_2 + \vec{k}_3)
        \frac{6}{5} \fnl \sum_{i < j} P_\zeta(k_i) P_\zeta(k_j) \,,
        \label{fnldef}
    \end{equation}
one gets
$\fnl$ as in Eq. (\ref{fnl}).  
($\fnl$ as defined in eq. (\ref{fnldef}) is 
the negative of $\fnl$ in
Refs. \cite{mald,seerylidsey}.)
The key difference between our expression for $\fnl$ and that in 
Ref. \cite{seeryetal} and other works is 
that our $\fnl$ is a function of $t$ rather
than $t_*$ because we expressed $\hat\zeta(t)$ as a function of $\hat Q(t)$
instead of $\hat Q(t_*)$.

Our calculation of
$\langle\zhat^3\rangle$ above has been rather straightforward.
We neither use extensive integration by parts to rewrite
the action nor do we invoke any field redefinition
as in Refs. \cite{mald,seerylidsey,chenetal}.  
Furthermore, if $t$ corresponds to the end of inflation 
then $N_e\approx60$ for modes entering the horizon
today.  
Then, as mentioned earlier, the $\xi\Ne$ term in $\fnl$ is comparable to
the $\eta$ term and a priori should not be ignored. 

Now we consider the
constancy of the 3-point function of $\hat\zeta$.
To determine how $\fnl$ changes during inflation 
after a mode crosses the horizon,
we take the derivative of $\fnl$ with respect to 
$\t$.
Now using
$d/dt = \dot\phi\,d/d\phi$
we get 
\begin{eqnarray}
         \frac{d\epsilon}{d t}
        &\simeq& [4\epsilon^2 -2\eta\epsilon]\,H
\nonumber\\
         \frac{d \eta}{d t} &\simeq&
        [2\epsilon\eta - \xi]\,H
\nonumber\\
        \frac{d\xi}{d t} &\simeq&
        [4\epsilon\xi - \eta\xi - \sigma]\,H
        \label{SRevol}
    \end{eqnarray}
where $\sigma=V^{\prime 2}V^{\prime\prime\prime\prime}/V^3$.
Then
$d\fnl/d\t\approx (5/6)\,d[-\xi \Ne-\eta]/d\t
=0$.
Here we have kept terms to first order
in slow roll parameters as in $\fnl$.  
Thus for the terms considered above
$d\fnl/dt$ is zero indicating
that $\fnl(\t)=\fnl(t_{\rm{ex}})$, i.e.,
the $\Ne$ factor in $\xi\Ne$, which corresponds to
growth outside the horizon,
is cancelled by changes in other terms
and does not contribute to the bispectrum.
\footnote{Note that these arguments are for adiabatic fluctuations of the
inflaton and
do not apply to any non-adiabatic fluctuations or fluctuations of
other scalar fields considered in
Refs.
\cite{falketal,zald,bern}.}
However note that $\langle Q^3\rangle$ in Eq. (\ref{Q3}) is a function of
terms proportional to $\xi$ and $\epsilon$ and so the $N_e$ dependent term
in $\langle Q^3\rangle$ is not cancelled by the time variation of other terms.

The argument above for the constancy of $\zeta$
closely follows that in Ref. \cite{seeryetal}.
In Ref. \cite{seeryetal} it is argued that $\fnl$ is constant 
outside the horizon
by taking the derivative of  $\fnl(t_*)$ effectively with
respect to $t_*$.  
One may argue that to study evolution of
$\langle \hat\zeta^3(t)\rangle$ outside the horizon one should
take derivatives with respect to $t$.  
But then $d\fnl/d\t$ will be 0 by construct, as
in the $\delta N$ formalism
$\hat\zeta(k,\t)$ is expressed in terms of $\hat\deltaphi(k,\tex)$
and $\fnl$ is independent of $\t$.

In conclusion,
we have calculated the 3-point function of
the curvature perturbation $\hat{\zeta}$ using the in-in formalism
in a gauge in which $\delta\phi\neq0$.
The calculation of the 3-point function of $\hat\zeta$
has generally been done ignoring inflaton self interactions,
and using the $\delta N$ formalism which by construct does not allow one to
check for evolution outside the horizon.
We have included the contribution
associated with non-gaussian
fluctuations of the inflaton due to
a cubic self interaction,
and this is proportional to $\xi\Ne$ and grows outside the horizon.
If we take $\Ne$ to be the number of
e-foldings of inflation after the mode of interest has left the horizon
till the end of inflation, then for our current horizon scale $\Ne$ is
60.  In new inflation, small field natural inflation and running mass models
of inflation,
$\epsilon<\xi<\eta$,
and $\xi\Ne$ is then comparable to other
contributions to the non-gaussianity parameter $\fnl$, and should not
be outrightly ignored.
The $\Ne$ dependent term corresponds to evolution outside the horizon.
However on including the time dependence
of other contributions to $\fnl$ this time dependent growth cancels.
Our results also indicate that there may be issues
related to the convergence at higher orders in perturbation theory. 
Higher order contributions can render $\fnl$ to have stronger time dependence.

\vskip 1cm
{\bf Acknowledgements} R.R. would like to acknowledge J. Maldacena and D.
Seery for very useful clarifications of their work.
R.R. would also like
to thank the organisers of the Xth Workshop on High Energy Physics
Phenomenology (WHEPP-X) at the Institute for Mathematical Sciences,
Chennai for discussions on issues related to this work.

\vskip 1cm
{\bf Appendix}

In this Appendix we compare the $\phidot(\tau)$,
$\phidot(\tau)^\prime$,$\phidot(\tau)^{\prime\prime}$,
$\phidot(\tau)^{\prime\prime\prime}$ 
and $\phidot(\tau)^{\prime\prime\prime\prime}$ terms 
in Eq. (\ref{I1_1}).
We will ignore terms proportional to $\epsilon$ and
derivatives of $\epsilon$ which are small for inflation models
that are of our interest.  We will also use
Eqs. (\ref{epsilon},\ref{eta},\ref{SRevol}).
Then 
$\eta\approx -\ddot\phi/(H\phidot)$, $\dot H= - \epsilon H^2\approx 0$,
and $\dot\eta\approx - \xi H$.  We take $k_i\sim k_t$.

The $\phidot$ terms within curly brackets
in Eq. (\ref{I1_1}) are $\sim \phidot/k_t$.

The $\phidot^\prime$ term is $\sim$ 
\be
\frac{k_1 \tau }{k_t^2} \phidot(\tau)^\prime
=\frac{k_1 \tau }{k_t^2} a \ddot\phi
=-\frac{k_1}{k_t^2} \frac{\ddot\phi}{H}
\approx\frac{1}{k_t}\phidot\eta \, .
\ee
Thus this is smaller than the $\phidot$ terms.

The $\phidot^{\prime\prime}$ terms are $\sim \phidot^{\prime\prime}/k_t^3=a^2\tdotphi/k_t^3$.
Now $\ddot\phi\approx -H\eta\phidot$.  Therefore
$\tdotphi\approx\xi\phidot H^2 + \eta^2\phidot H^2$.
Then the $\phidot^{\prime\prime}$ terms are $\sim$
\begin{eqnarray}
&&\frac{a^2}{k_t^3}(\xi H^2+\eta^2 H^2)\phidot
\\
&&=\frac{a^2}{a_{\rm{ex}}^2 H_{\rm{ex}}^2}
(\xi H^2+\eta^2 H^2)\frac{\phidot}{k_t}\cr
&&=\exp(2N_e)(\xi+\eta^2)\frac{\phidot}{k_t}
\end{eqnarray}
where we use $k_t=a_{\rm{ex}} H_{\rm{ex}}$ and
ignore the variation in $H$. 
With $\xi= 0.5\,\eta^2$, $\eta=-0.02$ and $N_e=60$, the
above is approximately $10^{49}$ times
larger than the $\phidot$ term.

The $\phidot^{\prime\prime\prime}$ term is 
$\sim$
\be
\frac{k_1\tau}{k_t^4}\phidot^{\prime\prime\prime}
=\frac{k_1\tau}{k_t^4}a^3\fdotphi
=-\frac{k_1 a^2}{H k_t^4}\fdotphi
\ee 
Now $\dddot\phi\approx (\xi+\eta^2)H^2\phidot$.  Therefore
$\fdotphi\approx -(4\xi + \eta^2) \eta H^3\phidot$.  Then
the $\phidot^{\prime\prime\prime}$ term is 
$\sim$
\be
\frac{a^2}{a^2_{\rm{ex}} H^2_{\rm{ex}}} 
(4\xi+\eta^2)\eta H^2 \frac{\phidot}{k_t}
=
e^{2N_e}(4\xi+\eta^2)\eta\frac{\phidot}{k_t}
\ee
which is a factor of $\eta$ less than the $\phidot^{\prime\prime}$ term.

The $\phidot^{\prime\prime\prime\prime}$ terms are $\sim
\phidot^{\prime\prime\prime\prime}/k_t^5$.  Using 
$\fdotphi$ from above, $\fvdotphi\approx
(11\xi\eta^2+4\xi^2+\eta^4)H^4\phidot$.  Then the
$\phidot^{\prime\prime\prime\prime}$ terms are $\sim$
\be
\frac{a^4}{k_t^5}\fvdotphi
\sim 10\frac{a^4\eta^4 H^4}{a^4_{\rm ex}H^4_{\rm ex}} \frac{\phidot}{k_t}
=10 \,e^{4N_e} \eta^4 \frac{\phidot}{k_t}
\ee
For $\eta=-0.02$ and $N_e=60$, the above term is a factor of
$10^{98}$ larger than the $\phidot$ term.  It is also larger than
the $\phidot^{\prime\prime}$ term by a factor of $10\,\eta^2\exp{(2N_e)}$.

Thus terms higher in slow roll parameters in evaluating the integral
in $I_1$ are larger than the lowest order terms.  This also holds for
$I_2$ and $I_3$.

%%%%%%%%%%%%%%%%%%%%%%%%%%%%%%%%%

\end{document}